\documentclass[aps,twocolumn,byrevtex,showpacs,preprintnumbers]{revtex4}

\usepackage{SIunits}
\usepackage{graphicx}
\usepackage{hyperref}

\def\uthigh{319.9}
\def\utmedium{315.7}
\def\utlow{310.2}
\begin{document}

\title{Memory effect in triglycine sulfate induced by a transverse electric field: specific heat measurement}

\author{M C Gallardo}
\email{mcgallar@us.es}
\author{J M Mart{\'\i}n-Olalla}
\author{F J Romero}
\author{J del Cerro}
\affiliation{Departamento de F{\'\i}sica de la Materia Condensada. Instituto Ciencia Materiales Sevilla. Universidad de Sevilla-CSIC. PO Box 1065.E41080 Sevilla.Spain}
\author{B Fugiel}
\affiliation{August Che\l kowski Institute of Physics, Silesian University, Uniwersytecka 4, PL-40-007 Katowice, Poland}
\date{December 10, 2008.}
\preprint{doi: 10.1088/0953-8984/21/2/025902; J. Phys.: Condens. Matter (2009) 21 025902}

\begin{abstract}
The influence of a transverse electric field in the specific heat of triglycine sulphate (TGS) has been studied. The specific heat of TGS has been measured heating the sample from ferroelectric to paraelectric phase after prolonged transverse electric field (i.e. perpendicular to the ferroelectric axis). It is shown that the specific heat of TGS can remember the temperature $T_s$ at which the transverse field was previously applied.
\end{abstract}

\maketitle

Triglycine sulfate $[(\mathrm{NH}_2\mathrm{CH}_2\mathrm{COOH})_3\cdot\mathrm{H}_2\mathrm{SO}_4]$ (TGS) has grown great importance for the development of infrared sensors\cite{hadni-jap-1969,white-jap-1964} and flat-panel displays\cite{rosenman-jap-1996} due to its large pyroelectric coefficient\cite{hoshino-pr-1957} and large pyroelectric figure of merit\cite{saturio-jjap-1995}. The crystal exhibits a ferroelectric phase (ferroelectric axis being parallel to axis $\mathbf{b}$) below $T_c= \unit{322}{\kelvin}$. At this temperature the substance shows a continuous phase transition to a paraelectric phase. 

 An electric field $E_\parallel$ parallel to the ferroelectric axis in TGS and its family has strong influence in the physical properties close to the critical point as revealed by numerous studies\cite{hoshino-pr-1957,taraskin-spss-1970,saturio-pss-1980,fugiel-prb-1990,westwanski-prb-1994,fernandez-prb-1998,javier-jpc-2005} and causes a single-domain (or almost single-domain) state. After field disconnection regions with opposite polarization randomly grows and the system tends to its equilibrium state for $E=0$. The longitudinal field has then a non permanent influence in the state of the system.

 The scenario is quite different for transverse electric field $E_\perp$ whose influence is reported to persist long time after switch-off although original properties can be easily restored after annealing the sample above the critical temperature. These properties are commonly named \emph{transverse field effect}. Among them, the following list: reduced or even vanishing hysteresis loop\cite{cwikiel-jpc-2000} (the reduction depends on how much time the field acted); the existence of temperature-dependent reversible transverse polarization\cite{fugiel-ssc-2002,fugiel-jpc-2002} and domain wall structure parallel to $\mathbf{c}$-axis\cite{cwikiel-physica-2000}. Dielectric measurements also showed a discontinuity in the slope of the inverse dielectric susceptibility\cite{fugiel-ssc-2006} at the maximum temperature of heating after transverse field application as well as a suppression of the low frequency dispersion\cite{fugiel-ssc-2006} observed from the temperature at which the transverse field was applied to the critical temperature. Such behaviour can be considered as some kind of \emph{memory effect}.
 
The transverse field effect in TGS has always been reported in directional properties but it would be of great interest to ascertain if these effects can be observed in bulk properties such as specific heat which would also be significant from the energetic point of view. It is also of interest for applications of this material since the specific heat plays a major role in determining the pyroelectric figure of merit.

In this letter, we present measurements of the specific heat \emph{after} prolonged transverse electric field and report the existence of persistent memory effect in the specific heat of TGS.

The specific heat measurements were performed using conduction calorimeter previously described\cite{jaime-jta-88,jaime-jsi-87}. We should quote its main features: able to provide specific heat absolute values, able to measure small changes of enthalpy (some milli joules) by means of high resolution differential thermal analysis (DTA) trace, able to measure in electric field, small rate of temperature change (as low as few decikelvin per hour). These features have been applied for observing ferroelectric phase transitions specially in the influence of electric field and for measuring latent heat even in the neighborhood of a tricritical point where it is expected to be comparatively small. \cite{jose-prb-1999,jose-jpc-2000,javier-jpc-2005}

\begin{figure}[t]
  \centering
  \includegraphics[width=6cm]{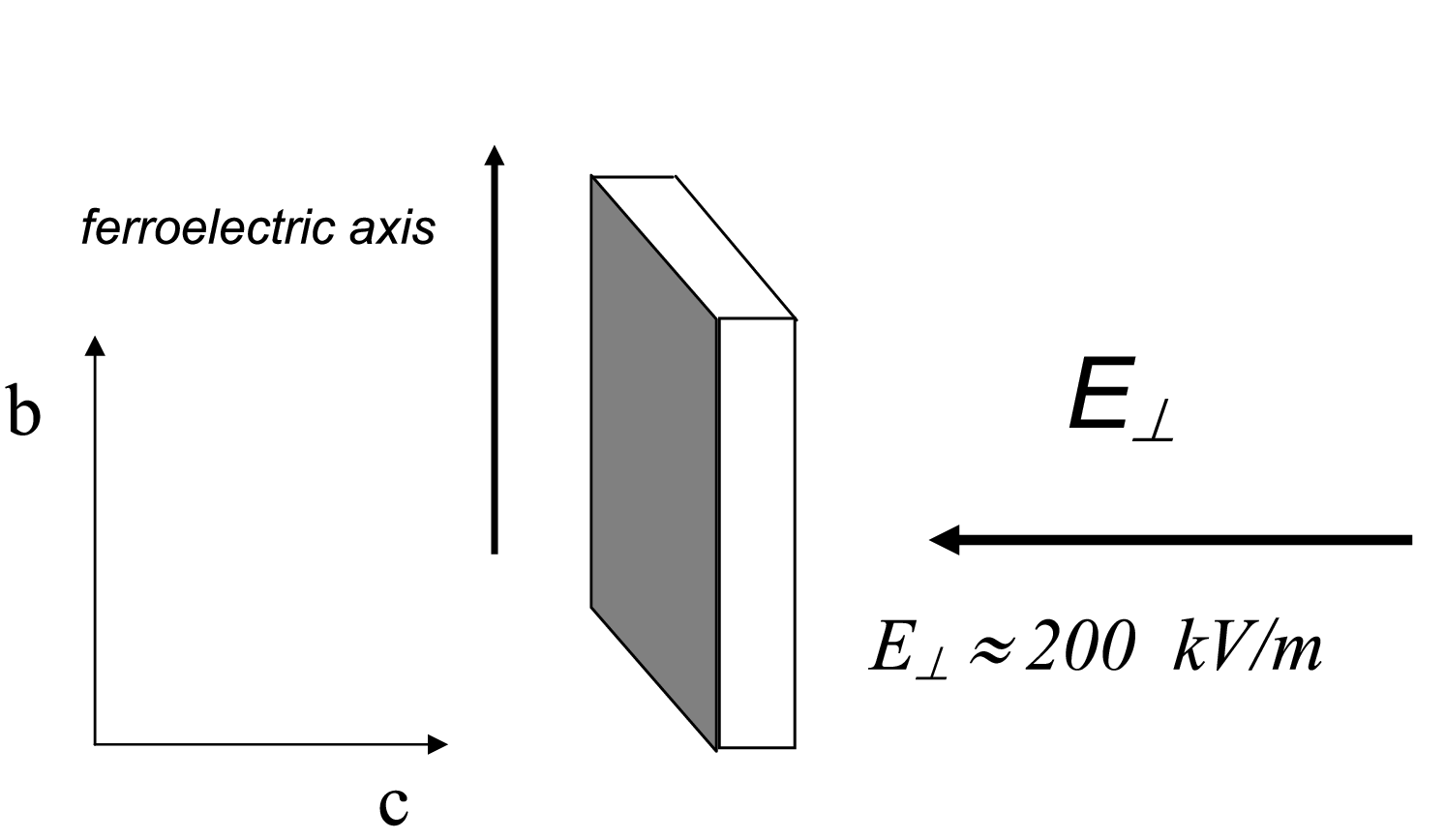}
  \caption{Sample geometry relative to electric field action. A silver solution (gray) was painted on faces perpendicular to $\mathbf{c}$ axis.}
  \label{fig:sample}
\end{figure}

The TGS crystal used in this experiment was produced by MolTech GmbH in Berlin. It was of the form of a rectangular parallelepiped with dimensions $\unit{7.0}{\milli\meter}\times \unit{7.0}{\milli\meter}\times \unit{4.0}{\milli\meter}$ along $\mathbf{b}\times\mathbf{c},\mathbf{b},\mathbf{c}$ and $\unit{302}{\milli\gram}$ in mass. The electric field ($\unit{2}{\kilo\volt\usk\centi\reciprocal\meter}$) was applied perpendicular to the ferroelectric axis $\mathbf{b}$ and parallel to $\mathbf{c}$ (see figure~\ref{fig:sample}).  Sides perpendicular to $\mathbf{c}$ were previously painted with a thin layer of a silver solution as an electrode. 

The following sequence was kept in our experiments: the sample was cooled from the paraelectric phase down to a given temperature $T_s$ in the ferroelectric phase where the transverse electric field ($\unit{2}{\kilo\volt\usk\centi\reciprocal\meter}$) was applied for a time $t$ (typically $\unit{100}{\hour}$) and then disconnected; the sample was cooled below $\unit{298}{\kelvin}$ and then heated above critical temperature at a rate of $\unit{0.2}{\kelvin\usk\reciprocal\hour}$ while recording specific heat data. As a reference we also recorded specific heat data in a heating run after annealing in the paraelectric phase and without any kind of exposure to transverse electric field.

\begin{figure}[t]
  \centering
  \includegraphics[width=8.5cm]{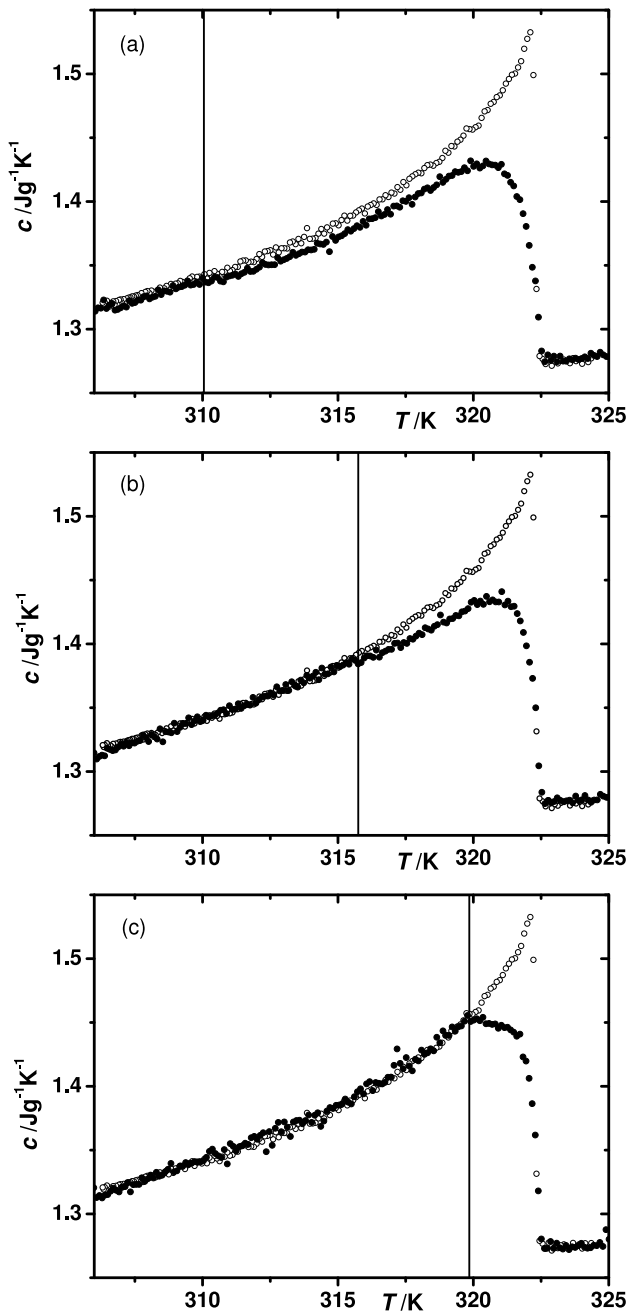}
\caption{\label{fig:datos}Memory effect in specific heat of TGS. Light circles show the reference ---no field--- specific heat for a heating run. Dark circles show the specific heat (heating run) after (1) applying transverse field $\unit{2}{\kilo\volt\usk\centi\reciprocal\meter}$ for $\unit{100}{\hour}$ at $T_s$ ---vertical line---, (2) field disconnection and (3) cooling down to $\unit{298}{\kelvin}$.}
\end{figure}

Figure~\ref{fig:datos} shows data for the experiments. Light circles always apply the reference experiment (heating run and no exposure to electric field). Dark circles in figure~\ref{fig:datos}(a) show data for $T_s=\unit{\utlow}{\kelvin}$ and $t=\unit{100}{\hour}$. Specific heat data in the ferroelectric phase differ from reference data above $T_s$ showing a memory effect. It is also shown that critical temperature does not noticeably change but the peak of the specific heat decreases. Figure~\ref{fig:datos}(b) shows data for $T_s=\unit{\utmedium}{\kelvin}$ and $t=\unit{100}{\hour}$ with analogous behavior. Figure~\ref{fig:datos}(c) shows data for $T_s=\unit{\uthigh}{\kelvin}$ and $t=\unit{100}{\hour}$, now somehow close to the critical point, also with analogous behavior. The last experiment was also run for $t=\unit{50}{\hour}$ and $t=\unit{10}{\hour}$ with same results, in agreement with Ref~\cite{kikuta-ferro-2006}. It is also noteworthy to mention the persistence of the phenomenon: the memory effect is observed above $T_s=\unit{\uthigh}{\kelvin}$, eleven days after field disconnection. 

\begin{figure}[t]
  \centering
  \includegraphics[width=8.5cm]{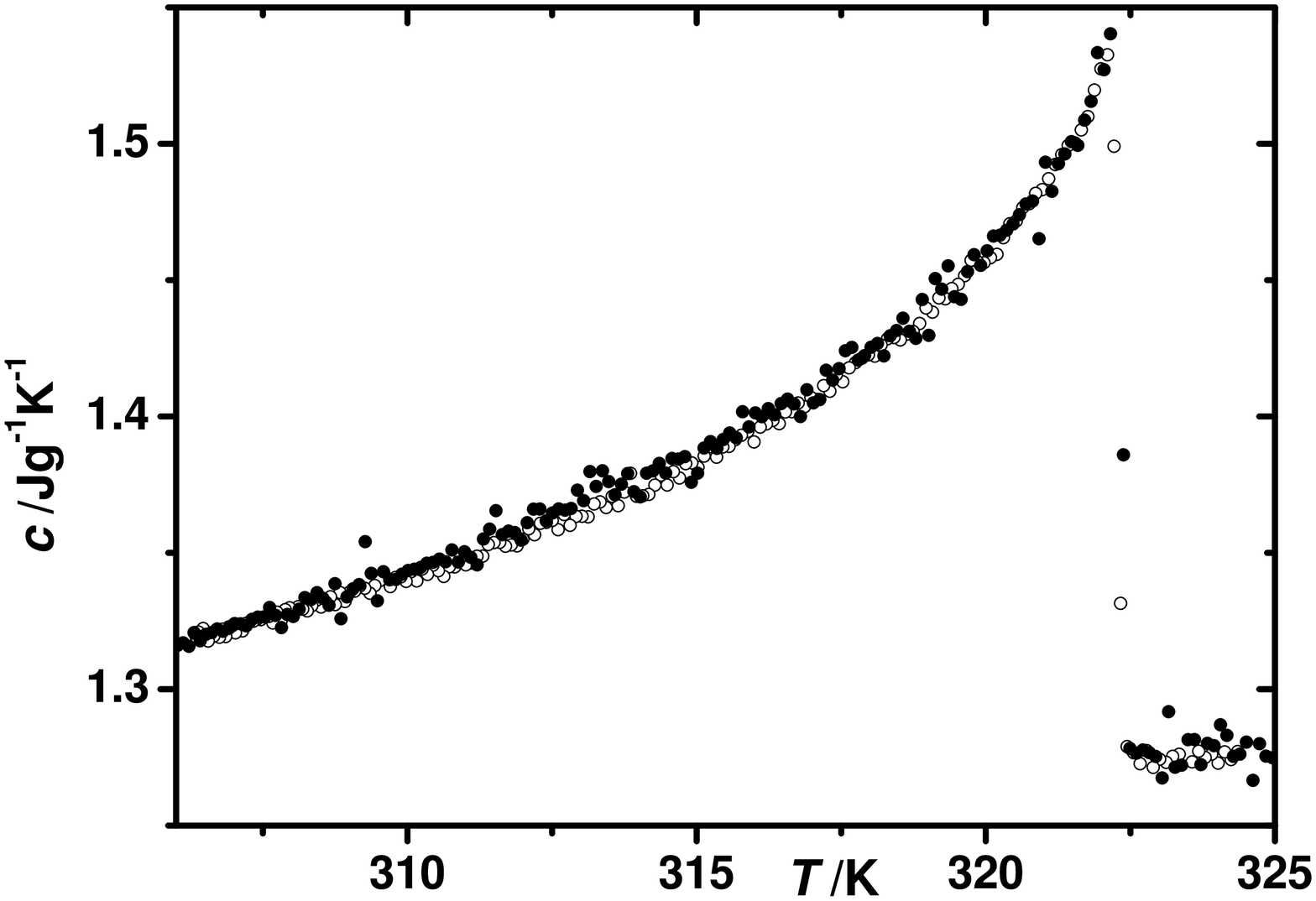}
  \caption{\label{fig:anneal}Rejuvenation of specific heat after annealing. Light points show the reference data in figure~\ref{fig:datos}. Dark points were taken after: (1) experiment in figure~\ref{fig:datos}(c) (dark points) , (2) annealing above $T_c$, (3) cooling to $T_s=\unit{\uthigh}{\kelvin}$, (4) holding temperature at $T_s$ for three days (no field applied), (5) cooling the sample to $\unit{298}{\kelvin}$ and (6) heating the sample while measuring specific heat (dark points).} 
\end{figure}

It should be finally stressed that the memory effect disappears in any case after rejuvenation in the paraelectric phase. Figure~\ref{fig:anneal} shows the reference data used in figure~\ref{fig:datos} and specific heat measurement after the sequence of experiments in figure~\ref{fig:datos} and annealing in the paraelectric phase. Both set of data now match to each other. Incidentally, figure~\ref{fig:datos} also shows this phenomenon as, for instance, the anomaly at $T=\unit{\utmedium}{\kelvin}$ (figure~\ref{fig:datos}(b)) is not observed again in figure~\ref{fig:datos}(a) or figure~\ref{fig:datos}(c). The chronological sequence of experiments in figure~\ref{fig:datos} was (b)-(a)-(c).

It was previously reported (\cite{fugiel-jpc-2002} and references therein)  that $E_\perp$ at $T<T_c$ alters permanently the dielectric properties of a TGS crystal. We now show that it also alters the thermal behavior of the sample. Having in mind all the results illustrating the so called  transverse field effect in TGS we should ponder which of the previously reported experimental data obtained after the transverse field disconnection at $T_s$ differ from original data (i.e. not influenced by $E_\perp$ or rejuvenated) only for $T>T_s$ and may eventually have influence on the specific heat. On the basis of previous investigations we know that each temperature increase from \unit{300}{\kelvin} up to $T_c$  ---after the transverse field disconnection at $T_s(<T_c)$--- is followed by a special kind of electric charge excitation just above $T_s$ as revealed, for instance, after warming of a short circuited (by electrometer) transversely polarized sample\cite{fugiel-ssc-2002}. In these experiments, for $T<T_s$, the transverse electric current density $J_\perp$ behaves as the pyroelectric current density flowing along the polar direction in conventional pyroelectric measurements resembling symmetry in cooling and heating. But above $T_s$ a second transverse electric current density is triggered. This component is characterized by an irretrievable outflow of free charge carriers from the sample and it is only observed in heating experiment; on the contrary, the sample can be cooled without observing any effect of this kind. It should be also mentioned that the influence of $dP_\parallel/dT$ on conventional specific heat measurements of TGS crystals was previously reported (see for instance Ref.~\cite{lashley-apl-2007} and references therein).

\begin{figure}[b]
  \centering
  \includegraphics[width=6cm]{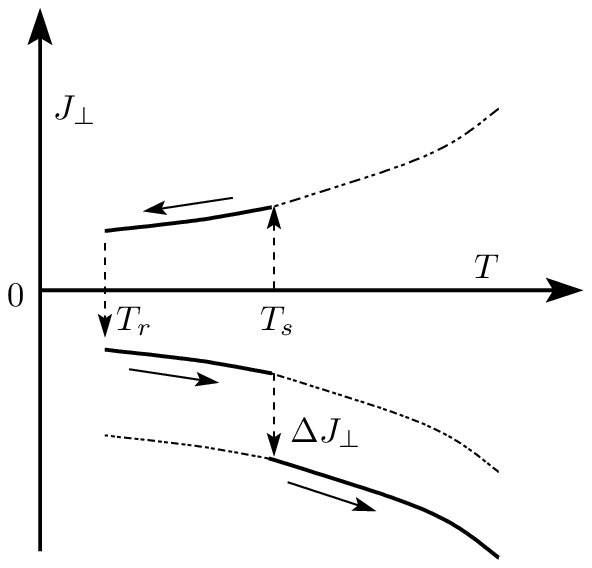}
  \caption{\label{fig:jout}Transverse current density $J_\perp$ (solid lines) in a TGS crystal after a transverse electric field disconnection at $T_s(<T_c)$ drawn schematically on the basis of Ref.\cite{fugiel-ssc-2002,fugiel-jpc-2002}. The sample is cooled from $T_s$ down to $T_r$ where the heating run begins. On approaching $T_s$ in heating an extra current is observed. Solid arrows indicate temperature change.}
\end{figure}

So during the zero-field specific heat measurements carried out after the transverse field action at $T_s$ (figure~\ref{fig:datos}) it is just while crossing $T_s$ that the process of releasing of depolarizing free charges begins. Such a process is triggered by a gradual decomposition of transversely polarized regions formerly \emph{frozen} by the prolonged transverse field applied at $T_s$\cite{fugiel-jpc-2002}. Although we deal here with an open electric circuit (a free-standing  sample with electrodes on the surfaces perpendicular to the c-axis), it is just at the temperature $T_s$ that the conditions of experiments change radically and the specific heat starts being influenced by $dP_\perp/dT$, more precisely by $\Delta J_\perp$ (see figure~\ref{fig:jout}). Although no electric current can flow during the specific heat measurement, the crystal can be then treated as a source of electric charges. 

Lashley et al\cite{lashley-apl-2007} analyzed specific heat on TGS crystals whose surfaces were either non-electrically grounded or short circuited (all of them) reporting strong influence of the grounding. We should stress that in the case of our data we are dealing with charges revealed from electrodes parallel to the ferroelectric axis. The surface perpendicular to the ferroelectric axis remains freestanding.

In summary, the existence of memory effect induced by transverse field in TGS has been shown in experiment concerning thermal (bulk) properties. The influence of the transverse filed is noticed long after its disconnection but only above the temperature $T_s$ at which the field was previously applied which is a memory effect.

This work has been supported by the Spanish Government, project FIS-2006-04045.


\end{document}